\documentstyle[11pt,IAU207_pasp,twoside]{article}
\input psfig
\def\edcomment#1{\iffalse\marginpar{\raggedright\sl#1\/}\else\relax\fi}
\reversemarginpar

\begin{document}
\title{3-Gyr-old Globular Clusters in the Merger Remnant NGC~1316:\ 
Implications for the Fate of Globular Clusters Formed During Gas-rich  
Galaxy Mergers} 
\author{Paul Goudfrooij}
\affil{Space Telescope Science Institute, 3700 San Martin Drive,
Baltimore, MD 21218, U.S.A.}

\begin{abstract}
The giant early-type merger remnant galaxy NGC~1316 is an ideal probe for
studying the long-term effects of a major merger on its globular
cluster (GC) system, given its spectroscopically derived merger age of
$\sim$\,3 Gyr which we reported in a recent paper. Here we report several
pieces of {\it photometric\/} evidence showing that the second-generation
GCs in NGC~1316 are at an evolutionary phase in between that of
luminous GCs found in younger merger remnants such as (e.g.) NGC 7252 and that
of `red' GCs found in `normal, old' ellipticals. The observation that
massive, second-generation GCs formed during major 
mergers can survive for at least 3 Gyr provides strong evidence that these
clusters can have `normal' mass functions including low-mass stars, and
hence that they can survive to reach `old age' similar to those of `normal'
ellipticals.  
\end{abstract}

\section{Introduction}

As this symposium emphasized once again, the study of globular clusters (GCs)
around galaxies provides unique and important information regarding our
understanding of how and when galaxies form. Their nature as simple stellar
population (SSP) significantly simplifies the determination of their ages and
metallicities relative to that of stellar populations that constitute the
integrated light of their parent galaxies. 
A particularly interesting and now well-known feature of the GC systems of
giant early-type galaxies is the presence of bimodal color distributions,
providing obvious evidence for the occurrence of a `second event' during the
formation history of these galaxies (as discussed by many authors in this
volume [e.g., Brodie, Kissler-Patig, Schweizer, Zepf]).  
To date, the only `simple' model that actually {\it predicted\/} such a
bimodality was that of Ashman \& Zepf (1992), which modeled the properties
of GCs forming in a major merger of gas-rich spirals (together with the 
giant elliptical host galaxies themselves). They figured that the enriched
gas associated with the spiral disks would form `red' (i.e., metal-rich) GCs
during the merger. Soon afterwards, observations of young merger remnant
galaxies with the {\it Hubble Space Telescope (HST)\/} revealed the presence
of luminous young GCs (e.g., Holtzman et al.\ 1992; Whitmore et al.\ 1993),
further supporting this `merger scenario'. 

However, certain properties of the GC systems around some giant ellipticals
(e.g., very large GC specific frequencies of 
central cluster galaxies) do not seem to fit into this simple merger scenario,
which prompted several groups to propose alternative concepts to explain
the GC color bimodality. One such concept is 
`multi-phase collapse' (Forbes, Brodie, \& Grillmair 1997) which
proposes that the metal-rich GCs are formed during a `secondary collapse'
phase of the galaxy which does not involve a merger. 
Another popular concept is the
`accretion scenario' (C\^ot\'e, Marzke \& West 1998) which proposes that
every galaxy is born with a GC system that has a median color according to
the color-magnitude relation among galaxies. The color bimodality in giant
ellipticals then arises through accretion of small galaxies by a large
galaxy. 

Given these different points of view, it is important to obtain additional
evidence for or against the `merger scenario'. Two rather major sources of
debate in this regard 
are:\ {\it (i)\/} Whether or not the GCs formed in mergers lack low-mass
stars so that they will disappear before reaching old age (see, e.g.,
Brodie et al.\ 1998; Smith \& Gallagher 2001), and {\it (ii)\/} 
whether or not general properties of GCs formed in mergers (i.e., their
spatial distributions, ages, metallicities and specific frequencies) are
compatible with those of `red' GCs in old  ellipticals. These two issues are
addressed in the remainder of this paper, using new results on the GC system
of the giant intermediate-age merger remnant galaxy NGC~1316. 

\section{Spectroscopic Evidence for 3-Gyr-old GCs in NGC~1316}

We (Goudfrooij et al.\ 2001, hereafter Paper~I) recently published results
of multi-slit spectroscopy of GC candidates in NGC~1316, an early-type galaxy
that is an obvious merger remnant (e.g., Schweizer 1980). We discovered the
presence of $\sim$\,10 GCs associated with NGC~1316 that have luminosities up
to $\sim$\,10 times higher than that of $\omega$\,Cen. Our measurements of
H$\alpha$ and the Ca\,{\sc ii} triplet in the spectra of the brightest GCs
showed them to have solar metallicity (to within 0.15 dex) and to have an age
of 3.0 $\pm$ 0.5 Gyr (Paper~I; also shown as Fig.~2b in Schweizer's
contribution to this volume). These are obviously GCs that have formed  
from enriched gas during a merger. This reinforces the view that luminous
GCs formed during mergers do not necessarily have high low-mass cutoffs to
their IMFs, and it also means that they can actually survive dynamical
disruption processes taking place during and after the merging process. 

\section{Color-Magnitude Diagram of the GC system of NGC~1316} 

We have studied the photometric properties of the GC system of NGC~1316
using a combination of {\it HST/WFPC2\/} and large-field ground-based 
photometry using the ESO NTT. The observations will be described in full
detail in a forthcoming MNRAS paper; here we only show and discuss relevant
results.  

\begin{figure}[th]
\centerline{
\psfig{figure=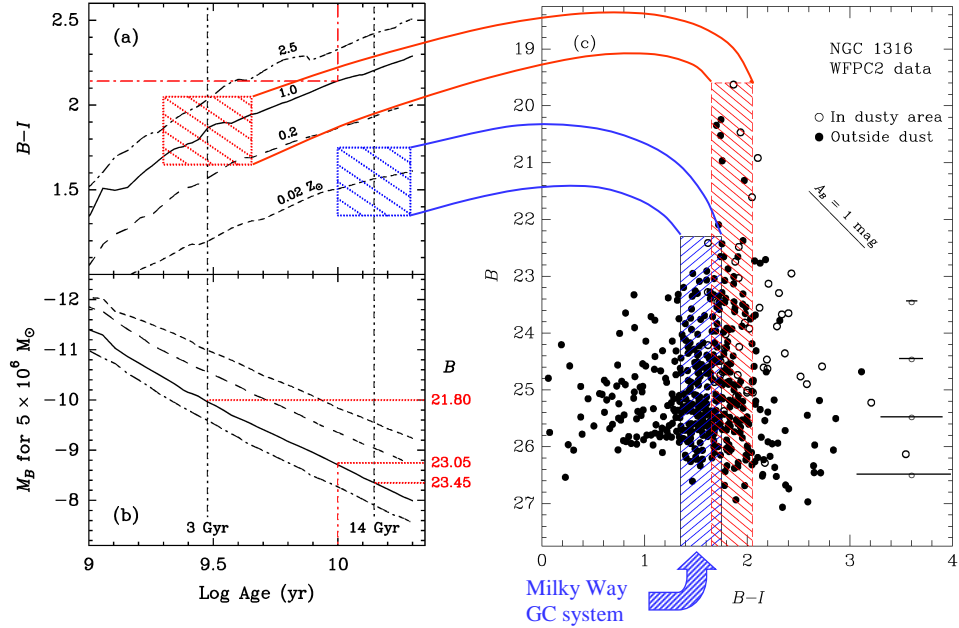,width=12.5cm,angle=-90.}
}
\caption[]{
 {\sl Panel (a):\/}\ Time evolution of $B\!-\!I$ according to the Bruzual \&\
 Charlot (1996) SSP models. Metallicities are indicated above the  models. 
 The hatched region on the {\it 
 left\/} indicates the $B\!-\!I$ color interval populated by GCs of age 3.0
 $\pm$ 0.5 Gyr and with [Z/Z$_{\odot}$] = 1.0 $\pm$ 0.15 dex, while
 the hatched region on the {\it right\/} does so for a old, metal-poor cluster
 population such as that of the Milky Way halo.
{\sl Panel (b):\/}\ Luminosity evolution for a GC having the mass of
 $\omega$\,Cen. Curves as in  Panel (a). $B$ magnitudes (at the distance of
 NGC~1316) for solar metallicity at 3, 10 and 14 Gyr are indicated on the
 right.   
{\sl Panel (c):\/}\ $B$ vs.\ $B\!-\!I$ CMD from the
 {\it HST\/} photometry. The two hatched regions are the same ones as
 indicated on Panel (a). Note that the hatched
 regions fit the observed CMD of NGC~1316 clusters very well.
}
\end{figure}

The $B$ vs.\ $B\!-\!I$ color-magnitude diagram (CMD) from the WFPC2 data is
shown in panel (c) of Fig.\ 1. The area which the GC system of the halo of
the Milky Way (Harris 1996) would occupy if placed at the distance of
NGC~1316 is also indicated (see arrow). 
The positions of GC candidates in this CMD are compared to predictions of SSP
synthesis models in panels (a) and (b) of Fig.\ 1 which show the time
evolution of 
$B\!-\!I$ and $M_B$ using the SSP models of Bruzual \& Charlot (1996, 
hereafter BC96). The metallicity range encompassed by the
brightest clusters at an age of 3~$\pm$~0.5 Gyr, $-\mbox{0.15} \la 
[Z/Z_{\odot}] \la \mbox{0.15}$, corresponds to $\mbox{1.75} \le B\!-\!I \le
\mbox{1.95}$ according to the BC96 models. Considering the typical 
photometric errors of the WFPC2 photometry as well, we assigned  
$\mbox{1.65} \le B\!-\!I \le \mbox{2.05}$ to the putative second-generation
population of clusters, indicated as the hatched region on the {\it 
left\/} side of panel (a). The other hatched region indicates
the $B\!-\!I$  interval occupied by the Galactic GC system 
which corresponds to a metallicity range of $-\mbox{1.7} \la [Z/Z_{\odot}]
\la -\mbox{0.8}$ (Harris 1996), which is consistent with the model
predictions at an age of 14 Gyr. Fig.\ 1 shows that these two 
intervals provide a very good fit to the positions of the 
cluster candidates on the CMD. Panel (b) shows that even after luminosity
dimming to an age of say 10 Gyr, the `red' clusters are up to several
magnitudes more luminous (and hence more massive) than that of the most
massive GC in our Galaxy. 

Our conclusion is that we are dealing with the presence of two distinct
cluster populations:\    
{\it (i)\/} a second-generation population clusters of roughly solar
metallicity that were formed $\sim$\,3 Gyr ago during a major spiral-spiral
merger, and 
{\it (ii)\/} a population of old, primarily metal-poor clusters that were
associated with the progenitor galaxies. 

\section{Radial Distributions of the GC (sub-)populations in NGC~1316}

The radial surface density distribution of GC candidates in NGC~1316 is
depicted in Fig.\ 2a, along with that of the integrated light of the galaxy.
The GC surface density is seen to flatten off somewhat towards the centre
with respect to the integrated light 
profile (note that this excludes the possibility of being due to extinction
by dust features in the inner regions). This is a known feature of GC systems
of `old' giant ellipticals, in which this flattening is more pronounced
(e.g., Forbes et al.\ 1998). It is most
probably due to the accumulative effects of tidal shocking (which is most
effective in the central regions) and dynamical friction during the first few
dynamical time-scales after the assembly of the galaxy (see, e.g.,
Fall and Vesperini's contributions to this volume). The more moderate
flattening of the surface density  
of the GC system of NGC~1316 towards the centre relative to those of
`old' ellipticals is consistent with the notion that NGC~1316 is only a
few Gyr old. As to the GC surface density profile in the outer regions,
fitting a power law results in a slope of $-1.36 \pm 0.10$, which is
consistent with the slope for the $B$--band light of the galaxy in the same
radial range ($-1.41 \pm 0.02$). This suggests that the GC system originally
experienced the same violent relaxation as did the main body of the galaxy. 

\begin{figure}[th]
\centerline{
\psfig{figure=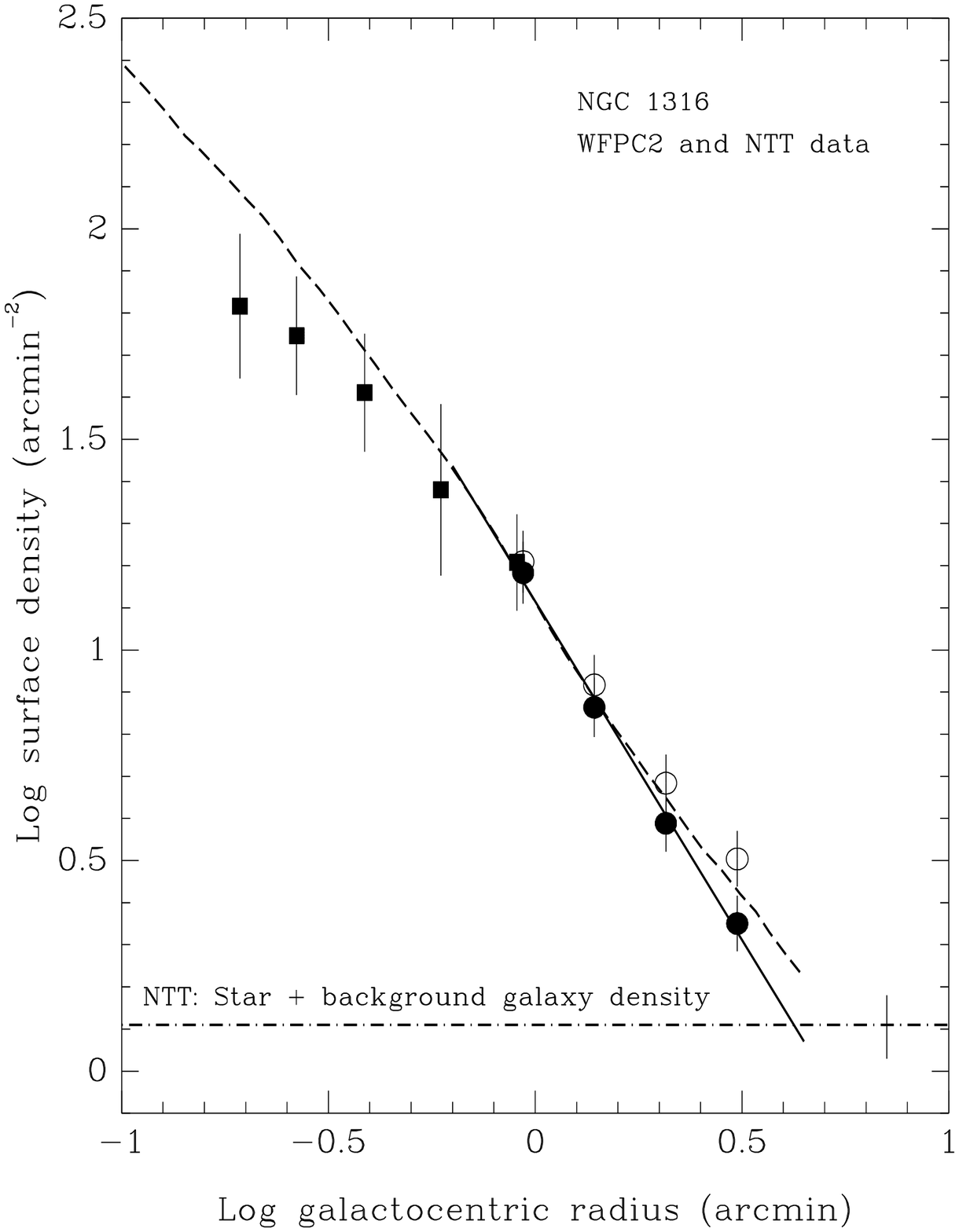,width=6.cm}
\hspace*{0.1mm}
\psfig{figure=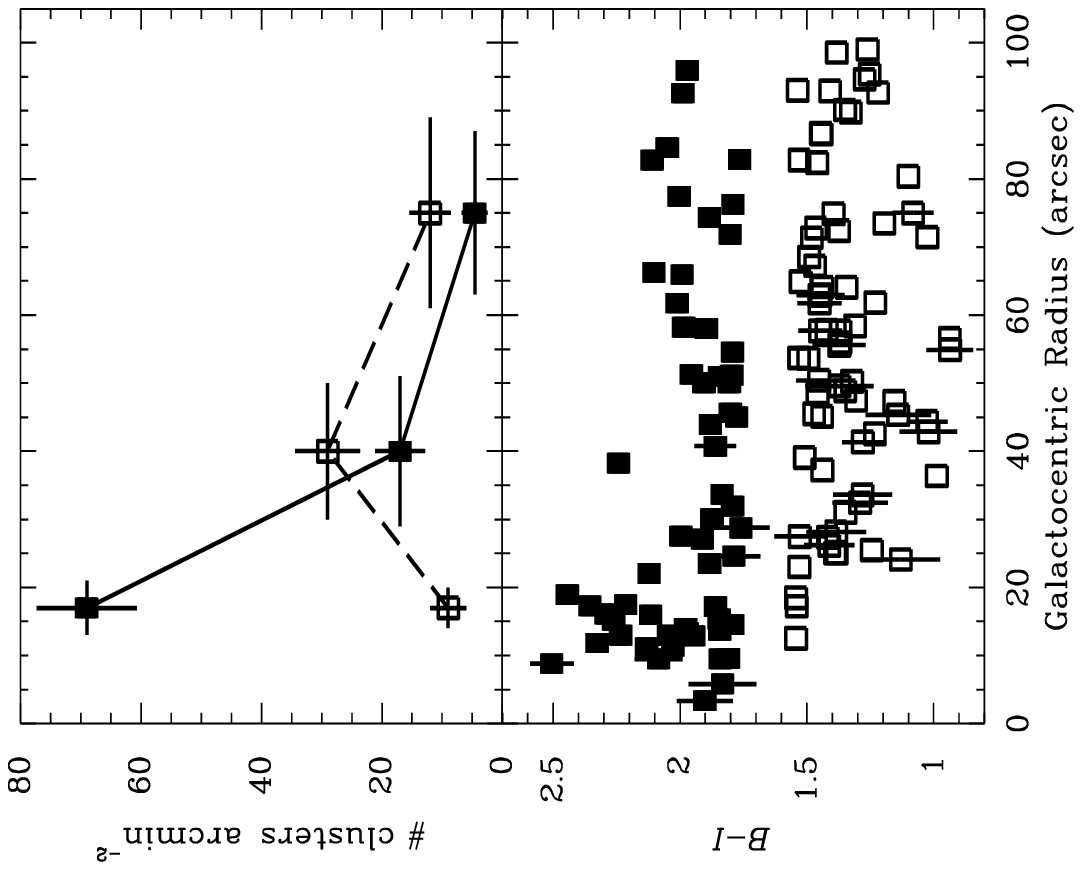,width=6.1cm,angle=-90.}
}
\caption[]{{\sl (a, left)\/} Radial distribution of the GC system of
 NGC~1316 down to $B$ = 24.0. Filled squares represent measurements from the
 WFPC2 data, corrected for foreground stars and background galaxies. Circles
 represent measurements from the ESO NTT data. The open and
 filled circles represent different corrections for the density of background
 galaxies. Reality is likely to lie in between the two sets of symbols. 
 A power-law fit to the solid circles beyond the inner 1 arcmin
 is shown by a solid line. The surface brightness profile of the integrated
 $B$-band light of the galaxy (normalized to the surface density of star
 clusters at a radius of 1 arcmin) is shown by the dashed line.
 {\sl (b, right)\/} Radial distributions of $B\!-\!I$ and surface density for
 the two GC subpopulations from the WFPC2 data. 
 Filled squares represent the `red' GCs and  
 open squares represent the `blue' GCs. }   
\end{figure}

A key prediction of the `merger scenario' of Ashman \& Zepf (1992) was that
the metal-rich GCs formed in mergers will be more centrally concentrated
than the pre-existing metal-poor GCs, which is the situation in well-studied
giant ellipticals (e.g,. Geisler, Lee, \& Kim 1996). NGC~1316 is an
important probe to test this prediction, since dynamical evolution has acted
over a much longer time scale in NGC~1316 than in merger
remnants studied to date (which are younger). To this end, we defined the
`blue' and `red' subpopulations as follows: 
$1.00 \leq B\!-\!I \le 1.55$ (`blue') and $1.75 \le B\!-\!I \le 2.50$
(`red'). This definition purposely avoids the area overlapping the two
hatched regions on Fig.\ 1c. The radial distributions of both subpopulations
are depicted on Fig.\ 2b. The `metal-rich' clusters are indeed clearly more
centrally concentrated than the `metal-poor' ones, consistent with the
predictions of the `merger scenario'. 

\medskip
Due to the page limit of this contribution, we refer the reader to a
forthcoming MNRAS paper (astro-ph/0107533) which will include other
photometric properties of the GC system of NGC~1316 (e.g., luminosity
functions for both subpopulations, the specific frequency, and its evolution). 

\medskip
\acknowledgments
It is a pleasure to thank the organizing committee (and Ms.\ Star Cluster and 
Doug Geisler in particular), who did a marvelous job in organizing this
Symposium. I'd also like to express a big {\bf thank you} to my excellent
collaborators on this project: Vicky Alonso, Markus Kissler-Patig, Jen Mack,
Claudia Maraston, Georges Meylan and Dante Minniti.






\end{document}